\documentclass[12pt]{article}
\usepackage{cite}
\usepackage{times,color,bm}
\usepackage{epsfig,amsmath,amssymb,amsfonts,mathrsfs,braket}
\usepackage{subfigure,graphicx,epstopdf}
\usepackage{tikz}
\usepackage{amssymb}
\usetikzlibrary{matrix,fit}
\usepackage[colorlinks]{hyperref}
\hypersetup{linkcolor = {blue},	citecolor = {blue},	urlcolor = {blue}}
\usepackage{textcomp}

\makeatletter

\newcommand{\Rmnum}[1]{\expandafter\@slowromancap\romannumeral #1@}
\makeatother

\newenvironment{sciabstract}{%
	\begin{quote} \bf}
	{\end{quote}}

\newcounter{lastnote}

\title{Quantum Advantage with Timestamp Membosonsampling}
\author
{Jun Gao,$^{1,2,\dagger}$ Xiao-Wei Wang,$^{1,2,\dagger}$ Wen-Hao Zhou,$^{1,2,\dagger}$ Zhi-Qiang Jiao,$^{1,2}$\\ Ruo-Jing Ren,$^{1,2}$ Yu-Xuan Fu,$^{1}$ Lu-Feng Qiao,$^{1,2}$ Xiao-Yun Xu,$^{1,2}$\\ Chao-Ni Zhang,$^{1,2}$ Xiao-Ling Pang,$^{1,2}$ Hang Li,$^{1,2}$ Yao Wang $^{1,2}$\\ Xian-Min Jin$^{1,2,*}$\\
	\normalsize{$^1$Center for Integrated Quantum Information Technologies (IQIT), School of Physics}\\
	\normalsize{and Astronomy and State Key Laboratory of Advanced Optical Communication Systems}\\
	\normalsize{and Networks, Shanghai Jiao Tong University, Shanghai 200240, China}\\
	\normalsize{$^2$CAS Center for Excellence and Synergetic Innovation Center in Quantum Information and}\\
	\normalsize{Quantum Physics,University of Science and Technology of China, Hefei, Anhui 230026, China}\\
	\normalsize{$^\dagger$These authors contributed equally to this work}\\
	\normalsize{$^\ast$E-mail: xianmin.jin@sjtu.edu.cn}\\
}
\date{}%\today

\begin{document}
	% Double-space the manuscript.
	\baselineskip24pt
	
	\maketitle
	
	\begin{sciabstract}
Quantum computer, harnessing quantum superposition to boost a parallel computational power, promises to outperform its classical counterparts and offer an exponentially increased scaling. The term ``quantum advantage" was proposed to mark the key point when people can solve a classically intractable problem by artificially controlling a quantum system in an unprecedented scale, even without error correction or known practical applications. Boson sampling, a problem about quantum evolutions of multi-photons on multimode photonic networks, as well as its variants, has been considered as a promising candidate to reach this milestone. However, the current photonic platforms suffer from the scaling problems, both in photon numbers and circuit modes. Here, we propose a new variant of the problem, timestamp membosonsampling, exploiting the timestamp information of single photons as free resources, and the scaling of the problem can be in principle extended to infinitely large. We experimentally verify the scheme on a self-looped photonic chip inspired by memristor, and obtain multi-photon registrations up to 56-fold in 750,000 modes with a Hilbert space up to $10^{254}$. Our work exhibits an integrated and cost-efficient shortcut stepping into the ``quantum advantage" regime in a photonic system far beyond previous scenarios, and provide a scalable and controllable platform for quantum information processing.		\\
	\end{sciabstract}	

	%\subsection*{Introduction}
During the last few decades, people have witnessed the flourish of quantum information science, a crossover between quantum mechanics and information theory\cite{Nielsen}. There have been several epoch-making theoretical progresses along with proof-of-principle demonstrations that involve quantum computing\cite{QC}, quantum cryptography\cite{GisinRMP}, and quantum error correction\cite{error}. Though the implementation of a large-scale universal and fully operational quantum computer is still beyond the capabilities of current technologies, there exists a growing sense of confidence and excitement that we are heading towards the era of noisy intermediate scale quantum\cite{NISQ}, especially when Google Sycamore processor\cite{supremacy} has demonstrated both the capability and scalability on the random circuit problem even not showing any practical applications. How to construct a controllable quantum system, and the system itself standing as a computationally hard problem meanwhile being able to be mapped to real-life applications, benchmark the regime of ``quantum advantage" in new era.

Boson sampling has become a prevailing candidate to demonstrate quantum advantage ever since Aaronson and Arkhipov formulated the computational problem comprehensively\cite{AA}. The races between boson sampling machines and supercomputers are being extensively boosted when many protocols targeting on real-life applications are proposed based on the kernel of boson sampling\cite{Vibronic,Spin,Graph1,Graph2,Graph3}. A typical boson sampling setup consists of the following parts: $n$ bosons interference in a linear network described by an $m$-mode Haar-random unitary matrix, and then are detected by on-off detectors. Relying on two unproven but plausible conjectures, it is argued that even approximating the probability distribution of a boson sampling machine seems classically intractable unless the polynomial hierarchy collapses. 

There have been constant endeavors to pursue large-scale boson sampling experiments, from the seminal prototypes\cite{Oxford_2013,Vienna&Jena_2013,Roma&Milano_2013,Roma&Milano_Birthday_2013,Brisbane_2013,Roma&Milano_validation_2014,Bristol_Haar_random_2014,Bristol_Universal_2015,Vienna&Jena_2015,Roma&Milano_2016,Brisbane_QD_2017,Timebin,USTC_time_bin_2017,USTC_QD_2017,USTC_Lossy_2018,Oxford2019,Collisionfree} to the most recent advance\cite{USTC2019} trying to approach the point where commercial laptops cannot defeat such a quantum system, as well as variants of boson sampling using different input states, such as scattershot boson sampling\cite{Roma&Milano_Scattershot_2015,USTC_Scattershot_2018} and Gaussian boson sampling\cite{Bristol_Gaussian_2019}. The most challenging bottleneck is the scalability issue towards genuine quantum advantage regime, which is set by both photon sources and linear network mode number following the permanent-of-Gaussians conjecture. 

In this work, we propose and experimentally demonstrate quantum advantage with timestamp membosonsampling, which efficiently scale up the boson sampling problem by fully exploiting the registration time information of single photons and the recurrent mode resource of a self-looped photonic chip inspired by memristor. Quantum interferences in either intra or inter-layer time section can be all recorded individually and analyzed globally. The scale of the timestamp membosonsampling can be in principle extended to infinitely large, as the schematic in FIG.1(a) shows. Each layer of time section contains $n$ photons and $m$ circuit modes. A repetition rate of $N$ expands the scale of the setup $N$ times. Probability for each event can be reconstructed from the timestamp information with only few counts rather than the accumulated data.

To avoid simply enlarge the scale by repeating each single run $N$ times on a photonic chip, additional inter-layer interaction should be introduced to bring the coherence to global space. The overall unitary matrix is merely the direct sum of each intra-layer matrix $\Lambda$, visualized as a diagonal block matrix
\begin{equation}
\bigoplus_{i=1}^{N} \mathbf{\Lambda}_{i}=\operatorname{diag}\left(\mathbf{\Lambda}_{1}, \mathbf{\Lambda}_{2}, \mathbf{\Lambda}_{3}, \ldots, \mathbf{\Lambda}_{N}\right).
\end{equation}

Inspired by the mechanism of memristor\cite{Mem}, where the device is historically influenced by the past evolution, we introduce accurate temporal overlaps between different layer of time sections. All the layers of time section are coherently connected via a self-looped architecture by matching the inter-layer time delay and periodic interval, that is why the scheme is referred as membosonsampling. We depict different loop-based architectures in FIG.1(b). There are three different loop functions that can be employed to artificially construct the overall unitary matrix, including passive loop, active loop and multiple loop. Passive loop means the photons can participate in the next run of sampling process, while the active loop contains a storage-based method\cite{Hybrid} that allows the photons ``jump" into any other later runs on demand. If the loop-based architecture further introduces multiple loops, the corresponding matrix is more connected and complex than the single loop architecture. 

As the memristor-like effect increases, the final graph structure of membosonsampling becomes a nearly fully connected graph, as shown in FIG.1(c). Each node in the graph represents an intra-layer matrix, and the edges indicate that the photons interfering in the current layer of time section can traverse different layers as long as the photons enter the loop channel. 

The most straightforward way is to construct the loop by connecting the output channel back to the input channel and compensate the time delay $\tau$ to match the inter-layer time delay and periodic interval, which can contribute the off-diagonal terms in the scattering matrix, as the timeline shown by FIG.1(a) inset. The transition probability from the former layer to the next one is denoted as $p$, then the off-diagonal terms in the matrix describe the transition process. Even though the values in each cascaded layer decay exponentially, as long as the values are nonzero, the inter-layer quantum interference can always occur. 

The generated scattering matrix $U$ can be described as,
\begin{equation}
U=\left[\begin{array}{ccccc}
\Lambda_{1} & p^{1} \Lambda_{2} & \ldots & p^{N-2} \Lambda_{N-1} & p^{N-1} \Lambda_{N} \\
p^{N-1} \Lambda_{1} & \Lambda_{2} & \ldots & p^{N-3} \Lambda_{N-1} & p^{N-2} \Lambda_{N} \\
\vdots & \vdots & \ddots & \vdots & \vdots\\
p^{2} \Lambda_{1} & p^{3} \Lambda_{2} & \ldots & \Lambda_{N-1} & p^{1} \Lambda_{N} \\
p^{1} \Lambda_{1} & p^{2} \Lambda_{2} & \ldots & p^{N-1} \Lambda_{N-1} & \Lambda_{N}
\end{array}\right]
\end{equation}
By choosing appropriate layer number $N$ and $n^{\prime}$-fold coincidence events, our scheme can always fulfill the permanent-of-Gaussians conjecture. At the same time, this scheme automatically forms a scattershot scheme, and this combination between the calculation of permanents and the enhancement factor $\left(\begin{array}{c}{Nm} \\ {n^{\prime}}\end{array}\right)$ further increases the computational complexity (see Supplemental Materials for more details). The arbitrariness of choosing $N$ allows us make full advantages of photon and mode resources accumulated in time domain, to extend the scale of boson sampling problem and retrieve large $n^{\prime}$-fold coincidence events stepping into the regime of ``quantum advantage".

Our experimental setup is schematically shown in FIG.2(a). The preparation of multi-photon states is based on spontaneous parametric down-conversion\cite{Beamlike} (more details are given in Methods). The generated photons are injected into a photonic chip fabricated by femtosecond laser direct writing\cite{Fab} (see Methods) via a polarization-maintaining V-groove fiber array. One output of the waveguides is connected back to the input channel, and the temporal delay is matched precisely the time defined by the repetition rate of the Ti: sapphire oscillator, namely 80MHz. Such a self-looped architecture allows the further inter-layer quantum interference after an intra-layer quantum interference. Along the timeline, this architecture can infinitely expand the scale of both photon and circuit mode numbers. The other 15 output channels are coupled to a single-mode V-groove fiber array and then connected to the array of APDs. All the electronic signals are fed to the time-of-flight module, which can record the timestamp information of photon events in each channel with a time resolution up to 64ps.

Then we characterize the photonic chip and reconstruct the Haar random matrix\cite{Cha}. The randomness is achieved by randomly changing the coupling width between adjacent waveguides, which introduces randomness in a continuous quantum walk evolution\cite{QW_1D,QW_2D}. After the characterization of the full device, we activate the temporal loop channel. Once the memristor-like effect works, photons interfering in the current layer will inevitably influence the next layer. We depict the scattering matrix of a 4-layer structure in FIG.2(b), note that the survival probability after 3 layers leads to the nonzero values of off-diagonal terms. Introducing active loop and multiple loop by adding fast on-chip modulation\cite{LN} and more self-looped channels will further randomize the values of matrix elements in the overall unitary operation.

We name our integrated membosonsampling machine ``Zhiyuan". We synchronize the injected heralded single photons, and then run the machine for nearly 74 hours without any discontinuity. All photon events and timestamp information are recorded by the time-of-flight module, which produce a stored data up to 9.45 Terabyte and demand a high big-data processing capacity with parallel computing. We verify the validity of our scheme by reconstructing the distributions of 2-layer structure, which expand to a scale of 8 photons scattering in a 30-mode interferometer. We process all the data to extract multi-photon coincidence events, and take 4-fold events in 2 cascaded layers as an example, see Methods for processing details. 

We obtain 12,180 counts for all detectable combinations of input from 3-hour data. As is shown in FIG.3(a), we compare the experimentally measured photon distributions with the  theoretically calculated distributions. The ring figures demonstrate distributions of parts of the input combinations, and the black bars are experimental results while the gray bars represent the theoretical calculations. We calculated the average fidelity of the measured distributions in the subspace, as $F=\sum_{i} \sqrt{s_{i} t_{i}}$, and the obtained value is up to 86.7\%.

We have also repeated this procedure to retrieve 6-fold events in 2 cascaded layers, and the final counting number is 755. The average count rate for each possible output combinations becomes quite low so we utilize the timestamp method\cite{timestamp} to reconstruct the event probabilities (see Methods and more details are explained in the Supplemental Materials). Without accumulating the full picture of the output distribution, we perform the validation test successfully with the limited set of events against distinguishable sampler\cite{Roma&Milano_validation_2014}, see Methods for more details. The results are shown in FIG.3(b) and the monotonically increasing trend verifies that the data are collected from a genuine boson sampler. The validation results show that the genuine quantum interferences between different temporal layers and the loop architecture is functioning.

As the distinct feature and advantage of our scheme, involving more temporal layers into the post-processing can dramatically increase the photon events and enlarge the Hilbert space dimensions to infinitely large. We study the dependence of layer number, photon number and coincidence count rates. As shown in FIG.4(a) and (b), we test the influence of layer number and photon number on the photon coincidence event rates respectively. In FIG.4(a), the selected photon number is fixed to 5 and we vary the $N$ number from 100 to 1,000, with 100 as step size. To explore a more general relationship, we select the first 10,000 files (7 hour data) and divide them into 10 equal parts. The mean value and standard deviation of the results are shown in the histogram with error bar and the coincidence events are plotted in logarithmic coordinate. These results suggest that we can efficiently retrieve more photon events if we increase the layer number. In FIG.4(b), we conduct the same operation with the layer number fixed as 1,000 and change the desired photon coincidence number $n^{\prime}$ from 5 to 10. The experimental result also shows an exponential dependence. 

With the time deviation calibrations (see Supplemental Materials for more details) and the aforementioned tests, we use 50,000 layers to extract 56-fold photons events (112-fold photons events if including the triggers). The equivalent system scale of our Zhiyuan machine corresponds to 200,000 photons scattershot in a 750,000-mode interferometer with a Hilbert space up to $10^{254}$. We obtain 17,248 events and exhibit part of the events in FIG.4(c). The upper figure shows the layer numbers of the registered triggers while the lower squares represent the output ports in each specific layer. The blue squares represent the signals locating at the same temporal layer with the triggers. The red squares show the signals with the temporal loops functioning, and the darker the color, the stronger the memristor-like effect. These results have reached the ``quantum advantage" regime of boson sampling problem with extremely large Hilbert space, which surpasses all the demonstrated scales to the present and is commonly considered as an intractable task to verify even for a supercomputer\cite{tianhe,Sunway}. An efficient way to verify or validate large-scale photon coincidence events remains an open question\cite{SupremacyAA}. With more efficient temporal loops and source brightness, our scheme also functions as a promising platform to scale up the Hilbert space, and may serve as a versatile tool for quantum simulation\cite{Chalabi2019}.

In summary, we propose and experimentally demonstrate quantum advantage with timestamp membosonsampling machine Zhiyuan. Inspired by memristor, we link and introduce quantum inference between different temporal layers, which leads to capacity to scale up the boson sampling problem large enough to go beyond the tractable level of classical supercomputers. For a very large number of layer structure, our scheme can still work in the sparse distribution region by retrieving the timestamp information. The 56-photon events in 750,000 modes with a Hilbert space up to $10^{254}$ have demonstrated boson sampling problem and formed an exponentially large computational space. 

Reaching the ``quantum advantage" regime itself does not mean the practical applications. However, demonstrating an integrated and cost-efficient photonic ``quantum advantage" with boson sampling may lead to a new era of quantum computing since we can now try to map the classically intractable problems into the corresponding boson sampling kernels in a programmable fashion according to the developed theories. Our scheme serves as a key path to reach NISQ regime to explore the nature and potential applications of quantum physics in an entirely new region. Beyond the scenario of quantum computing, we may have many more perspectives on how to fully take advantages of such a straightforward way of constructing large-scale quantum systems in the full frame of quantum sciences and technologies.\\
	
	\subsection*{Acknowledgments}
This research is supported by National Key R\&D Program of China (2017YFA0303700), National Natural Science Foundation of China (NSFC) (61734005, 11761141014, 11690033), Science and Technology Commission of Shanghai Municipality (STCSM) (15QA1402200, 16JC1400405, 17JC1400403), Shanghai Municipal Education Commission (16SG09, 2017-01-07-00-02-E00049), X.-M.J. acknowledges support from the National Young 1000 Talents Plan and support from Zhiyuan Innovative Research Center of Shanghai Jiao Tong University.

\subsection*{Data availability.}
The data that support the findings of this study are available from the corresponding author upon reasonable request.
\\

	\subsection*{Methods}

	\paragraph*{Preparation of quantum light source}
A femtosecond pulse centered at 780nm from a mode locked Ti:sapphire oscillator is frequency doubled in a lithium triborate (LBO) crystal. The generated 390nm ultraviolet pulses are split into two paths by a balanced plate beam splitter and in each path the laser successively passes type-II a collinear and a beamlike cut beta barium borate (BBO) crystal. The BBO crystals produce two pairs of correlated photons via spontaneous parametric down-conversion. The collinear phase matching correlated photons are separated by a polarizing beam splitter and filtered by a long pass filter. All the generated photons are filtered by 3nm band pass filters (BPF) and then coupled into polarization-maintaining (PM) optical fibers with proper polarization control to ensure the same polarization of different paths. Now, we have prepared 4 pairs of correlated photons and each pair offers one photon directly detected by the avalanche photodetectors (APD) to serve as the triggers. The HOM interference visibility between different pairs is measured to be 83\%, which can certainly reveal the bosonic bunching feature of photons. Temporal indistinguishability for non-classical interferences among the down converted photons is achieved by scanning temporal delay lines.
	
	\paragraph*{Fabrication of photonic chip}
We focus the 513nm femtosecond-laser into the photonic substrate (Corning Eagle XG) by a 50X objective lens (NA 0.55) and the pulses with a repetition of 1MHz permanently change the refractive index of the focal zone. Then the substrate is continuously moved by the high-precision 3-D air-bearing stages to inscribe the boson sampling circuit with a velocity of 15mm/s. Before the laser enters the objective, we reshape the pulse shape into a narrow fringe by a cylindrical lens. The narrow fringe is parallel to the waveguides and the power is fixed at 210nJ. The circuits is 170\textmu m below the top surface of substrate and the input/output port separations are set at 127\textmu m, which are matched with typically commercial V-groove fiber arrays. We introduce the randomness of the circuits by varying different coupling widths. The original coupling width is defined as 10\textmu m, then we alter the uniform coupling widths by a random number ranging from (-1, 1)\textmu m. The 5-cm long chip has 16 modes of 6 input ports that lies at central part of the chip.

	\paragraph*{Detailed procedure on extracting multi-photon events for timestamp membosonsampling}
We first calibrate the signal delays for all the channels. In order to make up for the electronic delay differences among the signal channels and trigger channels, we scan the delay time of each signal and trigger channel and monitor their coincidence counts. The coincidence time window is set as 2ns and the delay time is set ranging from -37.5ns to 37.5ns with a 0.5ns step. We set trigger 1 as the standard timestamp, and then derive the best delay time for all the other channels. After timestamp calibrations, the trigger photons reach the time-of-flight module at the same time and signal photons arrive 5ns latter than the corresponding triggers. The following steps are the processing details: 

Step I, we browse all the trigger channels and filter out the events with 2 triggers while the signal channels contain more than 2 photons within 21ns after the first trigger's arrival time. The signal channels are more sensitive to the noises (background noises or multiple emissions), thus we exclude any suspicious events. The time window 21ns is chosen because this time scale is large enough cover all photons in such two consecutive sections and will not include photons in the third section. 

Step II, we mark the section number of the filtered result and then, we shift the second section by 12.5ns. By post-processing with a 2ns coincidence time window, we can retrieve the four fold events and filter out noisy photons. 

Step III, we check the results to make sure all the signals occur latter than the triggers since this is physically constrained by the experimental setup. For example, if there are 2 photons in the first section with only 1 trigger, we regard this as trigger missing, so we abandon the event. After this step, 2 signal photons and 2 trigger photons are perfectly matched to the timestamp of each other.

Step IV, we renumber the channels to match the newly generated scattering matrix. We can reconstruct the output distribution by either counts or timestamp information.

	\paragraph*{Time-of-flight module and timestamp method}
Our 32-channel time-of-flight module is directly connected to the array of APDs and record the arrival timestamp of the electric signals with a time resolution up to 64ps. We use the time information $\tau_{i}$ of each trigger as the timestamp to mark the occurrence of multi-fold coincidence events and the high time resolution enables us to reconstruct probability distribution accurately. The timestamp is related to the reciprocal of count (${N_i}^{-1}$). Then we can reconstruct the distribution by the timestamp with equation $p_i=\frac{\tau_i^{-1}}{\sum_{i}\tau_i^{-1}}$ by the first occurrence time of events. For multiple layer events, the timestamp of the whole event is marked by the last arrival trigger signal. The time-of-flight module and timestamp method enables us to identify the function of the self-looped structure and to reconstruct distribution with fewer events.	

	\paragraph*{Validation of boson sampling data}
In our experiment, we use statistic method to validate that the experimental results are boson sampling events against distinguishable sampling events, called likelihood ratio test. This method can significantly distinguish observed data from other samplers, in our case, we test our data against theoretically calculated classical distributions. 

At the start, given an estimator as $L_{k}=p_{k}^{i n d} / q_{k}^{d i s}$. Here $p_{k}^{i n d}$ and $q_{k}^{d i s}$ are the corresponding probabilities related to the event from indistinguishable and distinguishable samplers. The estimator starts at point $C=0$, then update it according to following rules,
\begin{equation}
C=\left\{\begin{array}{lr}{C,} & {a_{1}<L_{k}<1 / a_{1}} \\ {C+1,} & {1 / a_{1} \leq L_{k}<a_{2}} \\ {C+2,} & {L_{k} \geq a_{2}} \\ {C-1,} & {1 / a_{2} \leq L_{k}<a_{1}} \\ {C-2,} & {L_{k} \leq 1 / a_{2}}\end{array}\right.
\end{equation}
In our experiment,  $a_{1}$ and $a_{2}$ are chosen as 0.9 and 1.5 respectively. The counter is updated using our experimentally reconstructed data using the timestamp method. After several tens of updates, the boson sampler and distinguishable sampler will separate apparently.
	
	\clearpage
	% If your reference list includes text notes as well as references, include the following line; otherwise, comment it out.
	%\renewcommand\refname{References and Notes}
	
\clearpage
	
	%\noindent {\bf Fig. 1.}
	\begin{figure}[htbp]
		\centering
		\includegraphics[width=1.0\linewidth]{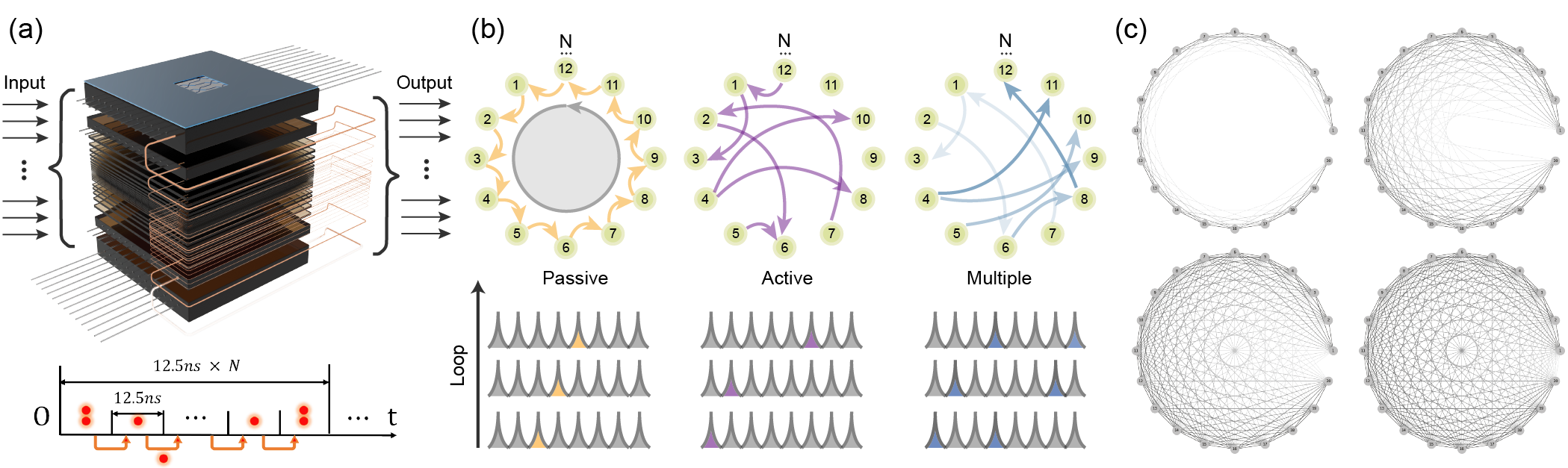}
		\caption{\textbf{Scheme of timestamp membosonsampling.} 
		\textbf{(a)} Each individual building block of the timestamp membosonsampling is described by the matrix $\Lambda$. The scale of the boson sampling problem can be extended to infinitely large using memristor-like effect and timestamp information. The timeline shows how the photons traverse to different temporal layers.
		\textbf{(b)} Prospective modification of loop architecture. Three different types of loop-based structure are depicted, namely, passive loop, active loop and multiple loop.
		\textbf{(c)} Graph structure of membosonsampling. The node represents each individual layer while the edges show the possible transitions among different layers. The more possible transition edges, the more complex the structure is.
	}
		\label{fig.1}
	\end{figure}
	
	\clearpage
	
	%\noindent {\bf Fig. 2.}
	\begin{figure}[htbp]
		\centering
		\includegraphics[width=0.95\linewidth]{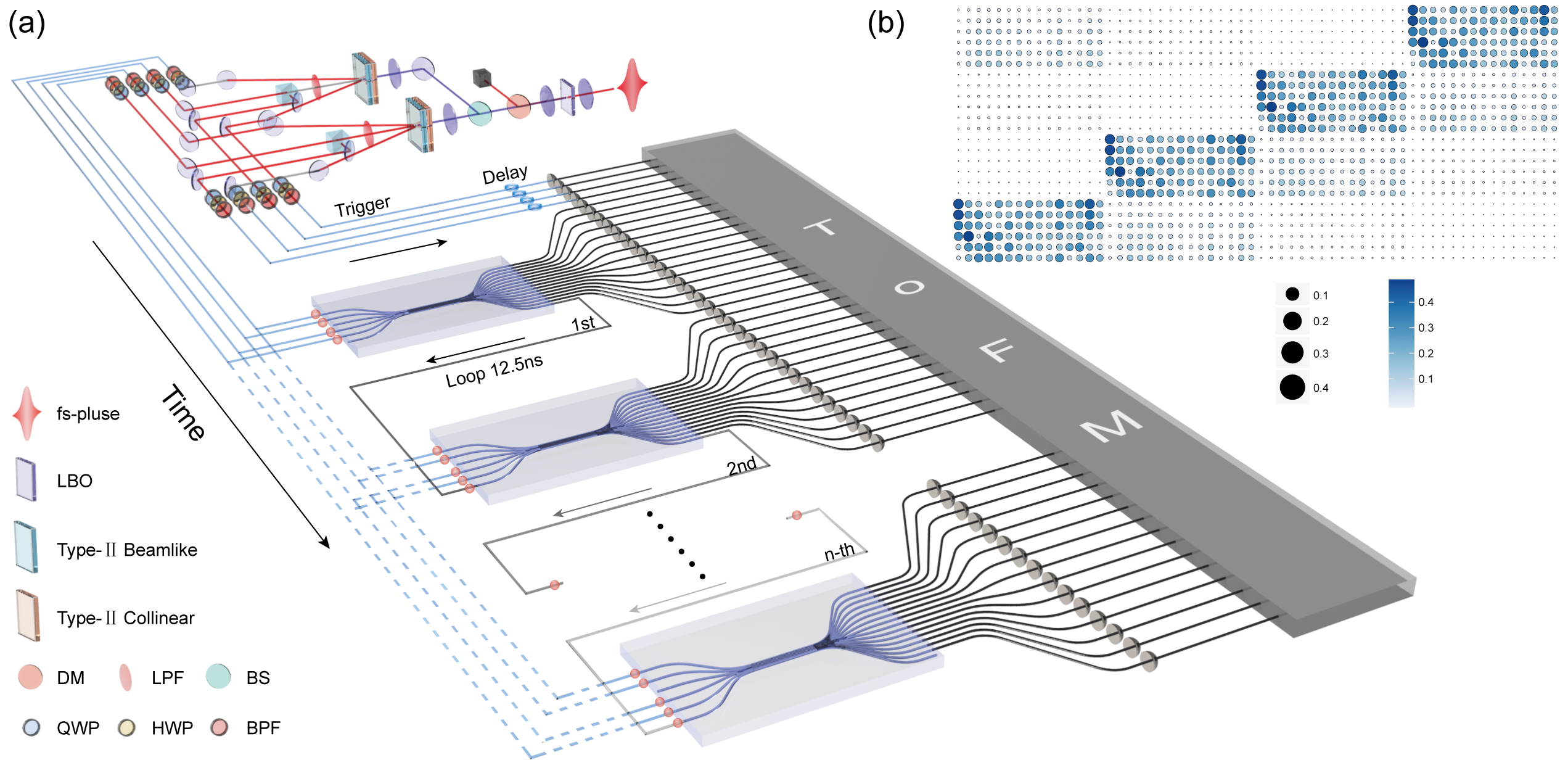}
		\caption{\textbf{Schematic of timestamp membosonsampling machine Zhiyuan.} 
		\textbf{(a)} The multi-photon Fock states are prepared by successively pumping BBO crystals with a frequency doubled 390nm femtosecond laser pulse. The heralded photons generated from down converted photon pairs are collected by fiber couples and are guided to the photonic chip. By taking each time interval as resources for both photons and circuit modes, the scale of the experiment can be extended $N$ times. The loop structure allows quantum interference between different temporal layers. The output in different layers are all addressed by a large array of APDs. The electronic signals are recorded by the time-of-flight module (TOFM), which can collect all the timestamps in the large Hilbert dimensions of the state space simultaneously. The timestamp information of the trigger channels heralds the corresponding photon coincidence events.
		\textbf{(b)} Experimental characterization of scattering matrix of a 4-layer structure. The survival probabilities are nonzero even after several cascaded layers.
	}
		\label{fig.2}
	\end{figure}
	
	\clearpage
	
	%\noindent {\bf Fig. 3.}
	\begin{figure}[htbp]
		\centering
		\includegraphics[width=1.0\linewidth]{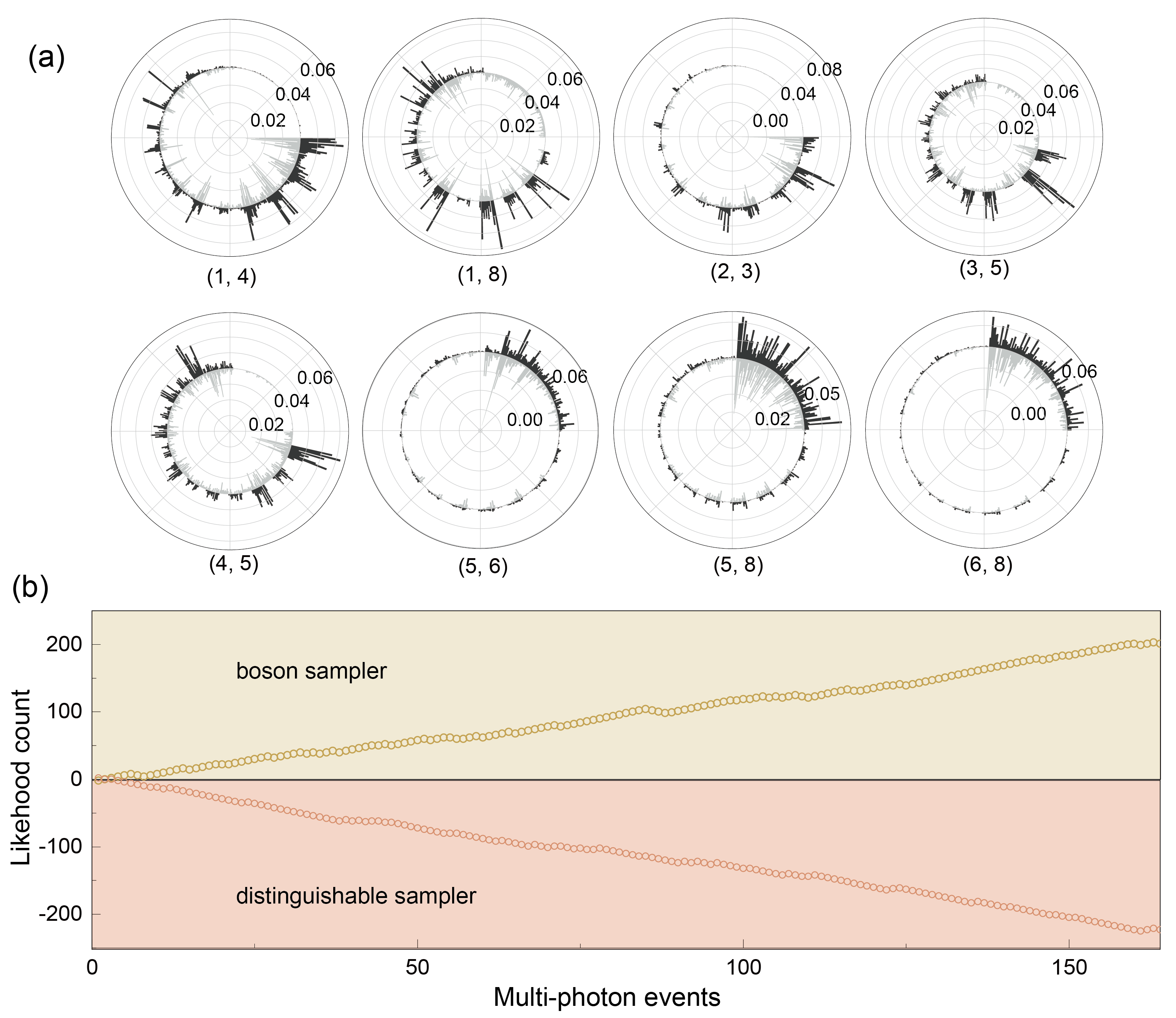}
		\caption{\textbf{Distribution and validation of membosonsampling experiments.} 
		\textbf{(a)} The comparison of 2-layer, 4-fold events probability distribution between the experimental data $p_i$ (dark grey) and theoretical data $t_i$ (light grey). The calculated fidelity $F$ is up to 86.7\%.
		\textbf{(b)} The validation of 2-layer, 6-fold event results. The experimental data are processed by the likelihood ratio test with only few events with the timestamp information. The experimental results (yellow dots) apparently deviate from the distinguishable sampler (red dots). The validation results reveal the occurrence of quantum interference between different temporal layers.
	}
		\label{fig.3}
	\end{figure}

	\clearpage
	
	\begin{figure}[htbp]
		\centering
		\includegraphics[width=1.0\linewidth]{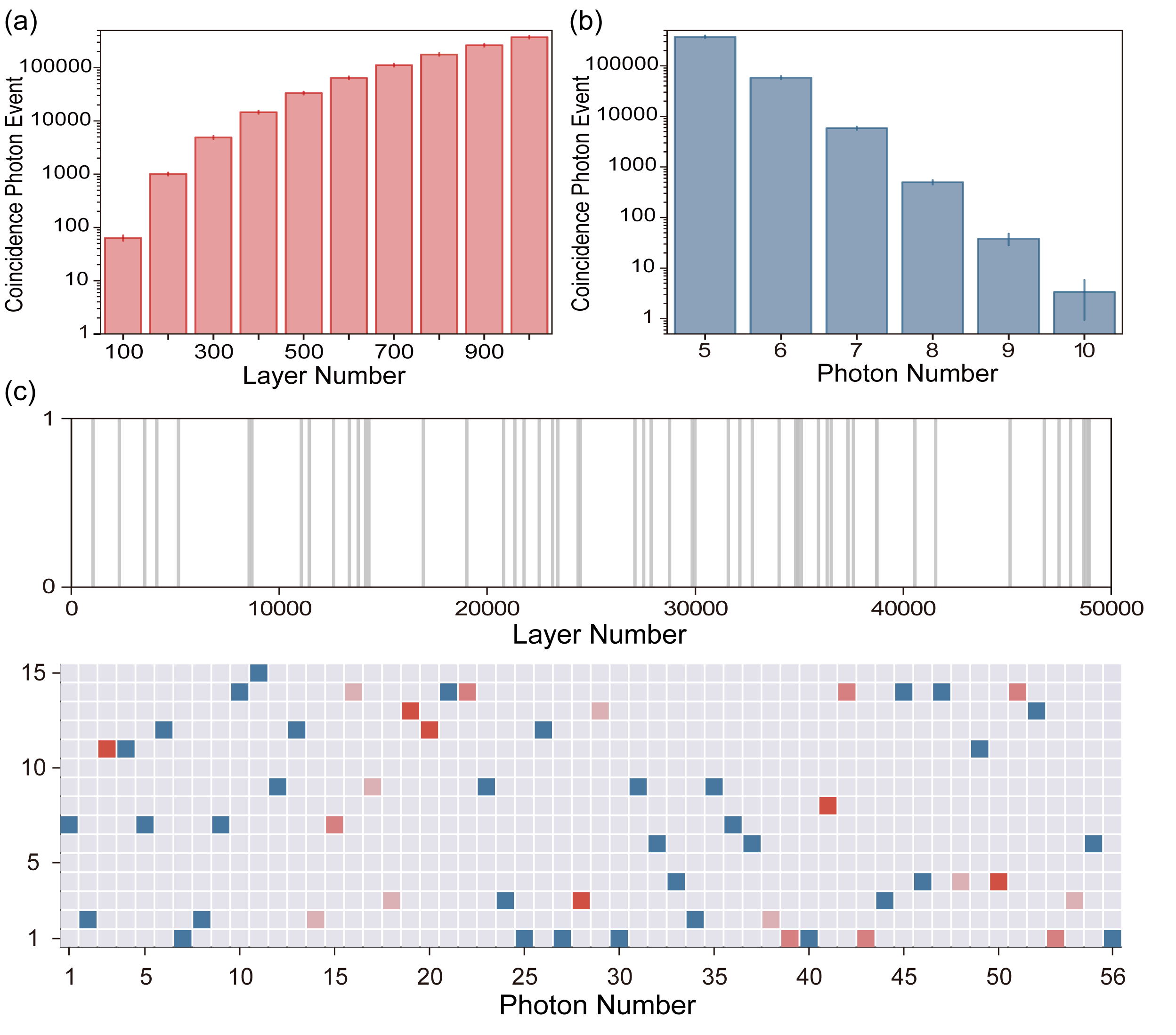}
		\caption{\textbf{Experimental results of multiple-layer data.} \textbf{(a)} Test of layer number $N$'s influence on coincidence photon events. We fix photon number to 5 and vary the layer number from 100 to 1000. The coincidence photon events are plotted in logarithmic axis. The increasing tendency exhibits the strong positive correlation between the layer numbers and coincidence events. 
		\textbf{(b)} Test of photon number's influence with layer number fixed to 1,000. The coincidence photon events drop with the increase of photon numbers. 
		\textbf{(c)} An example of a 56-fold photon event we have extracted. The upper figure is the distribution of the registered triggers' layer number. The lower figure exhibits the 56 output ports of the on-chip photons.
		}
		\label{fig.4}
	\end{figure}
	
	\clearpage
	
\section*{Supplementary Materials: Quantum Advantage with Timestamp Membosonsampling}

\noindent
{Jun Gao,$^{1,2,\dagger}$ Xiao-Wei Wang,$^{1,2,\dagger}$ Wen-Hao Zhou,$^{1,2,\dagger}$ Zhi-Qiang Jiao,$^{1,2}$\\ Ruo-Jing Ren,$^{1,2}$ Yu-Xuan Fu,$^{1}$ Lu-Feng Qiao,$^{1,2}$ Xiao-Yun Xu,$^{1,2}$\\ Chao-Ni Zhang,$^{1,2}$ Xiao-Ling Pang,$^{1,2}$ Hang Li,$^{1,2}$ Yao Wang $^{1,2}$\\ Xian-Min Jin$^{1,2,*}$\\
	% Shanghai
	\normalsize{$^1$Center for Integrated Quantum Information Technologies (IQIT), School of Physics}\\
	\normalsize{and Astronomy and State Key Laboratory of Advanced Optical Communication Systems}\\
	\normalsize{and Networks, Shanghai Jiao Tong University, Shanghai 200240, China}\\
	\normalsize{$^2$CAS Center for Excellence and Synergetic Innovation Center in Quantum Information and}\\
	\normalsize{Quantum Physics,University of Science and Technology of China, Hefei, Anhui 230026, China}\\
	\normalsize{$^\dagger$These authors contributed equally to this work}\\
	\normalsize{$^\ast$E-mail: xianmin.jin@sjtu.edu.cn}\\
}

	\baselineskip24pt
	
	\section{Enlarged graph structure of timestamp membosonsampling}
In FIG.S1, we depict the graph structure of membosonsampling with 20 layers. The nodes in the graph represent individual intra-layer matrix. The edges show the connections between different layers. Take node 1 as an example, photons in this layer can jump into any other layers. The probability of such transition process decays as the temporal distance becomes larger, as the color in the figure shows. The final graph structure is a nearly complete graph, which means in principle all the temporal sections can be connected. By taking enough temporal sections, our scheme owns the capability to extend the scale of boson sampling problem to extremely large, even into ``quantum advantage" regime.
	
	\begin{figure}[htbp]
		\centering
		\includegraphics[width=0.95\linewidth]{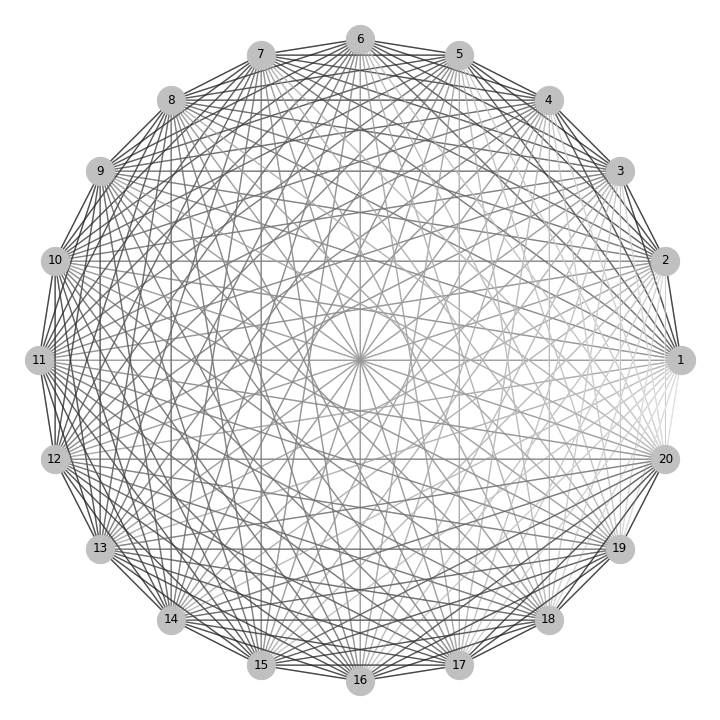}
		
		FIG. S1.\textbf{Graph structure of membosonsampling.}
		\label{s1}
	\end{figure}
	
	\clearpage
	
	\section{Computational complexity enhancement by timestamp membosonsampling}
By choosing an appropriate layer number $N$, our scheme can extend both the photon number and mode number by $N$ times. Thus if we choose the selected $n^{\prime}$-fold photon coincidence smaller than $Nn$, the experiment automatically turns into scattershot boson sampling. The possible input combinations scales as $\left(\begin{array}{c}{Nm} \\ {n^{\prime}}\end{array}\right)$, which also increases the computational complexity of the boson sampling problem. In FIG.S2, we vary the layer number $N$ and the $n^{\prime}$-fold coincidence, and calculate the complexity enhancement in logarithmic axis introduced by combinations. It is first argued by L. Latmiral \textit{et al.} that this combination factor can significantly enhance the consuming time to perform the exact classical calculation of full boson sampling distributions. We should also note that our scheme can directly reach the ``quantum advantage" regime by taking enough temporal sections. The photon number counts and the fulfilled permanent-of-Gaussians conjecture benchmark the capabilities of our experimental setup to reach NISQ regime.

	\begin{figure}[htbp]
		\centering
		\includegraphics[width=1\linewidth]{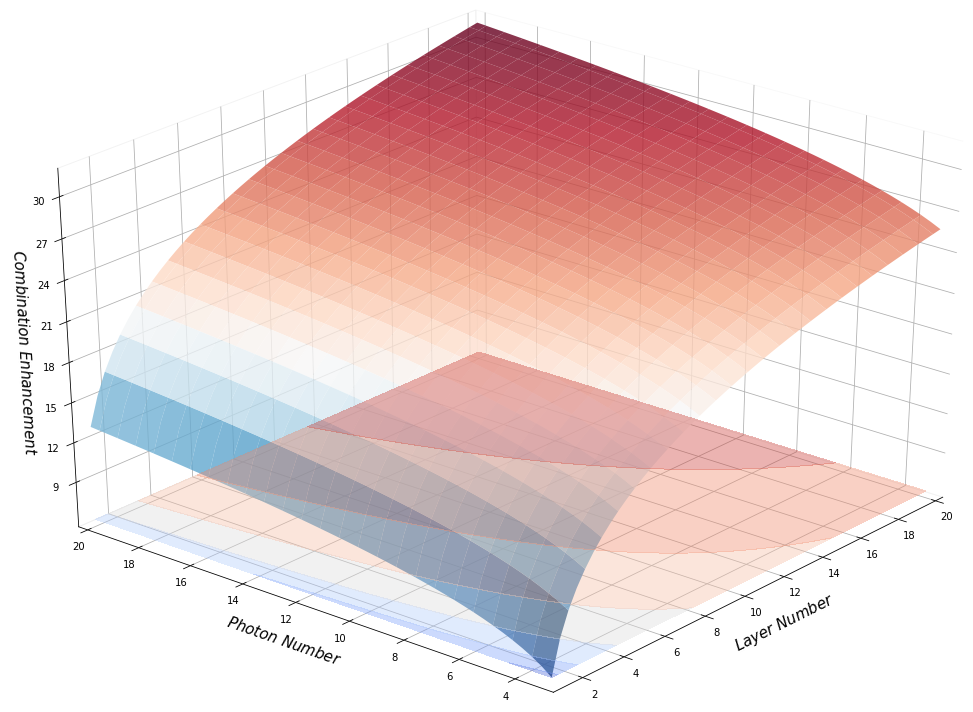}
		
		FIG. S2.\textbf{Computational complexity enhancement introduced by increasing input combinations.} Our scheme can automatically form a scattershot version of boson sampling. We vary the layer number $N$ and $n^{\prime}$-fold photon events to evaluate the computational complexity enhancement. The figure is depicted in logarithmic axis.
		\label{s2}
	\end{figure}
	
	\clearpage
	
	\section{Characterization of the boson sampling chip}
In essence, the boson sampling circuits can be represented by a unitary matrix, whose elements are complex numbers. The characterization process can be therefore summarized into two steps, calculating the real and imaginary part. In our experiment, we first inject heralded single photons into the input port and measure the intensity distributions of the output ports to calculate the moduli of each element. Then we use indistinguishable photons to scan the HOM interference curves and record all the coincidence counts of every two output ports. From these obtained HOM interference curves, we can determine the arguments and the signs of each element. The corresponding amplitudes and phases are shown in FIG.S3(a) and FIG.S3(b). 

	\begin{figure}[htbp]
		\centering
		\includegraphics[width=1\linewidth]{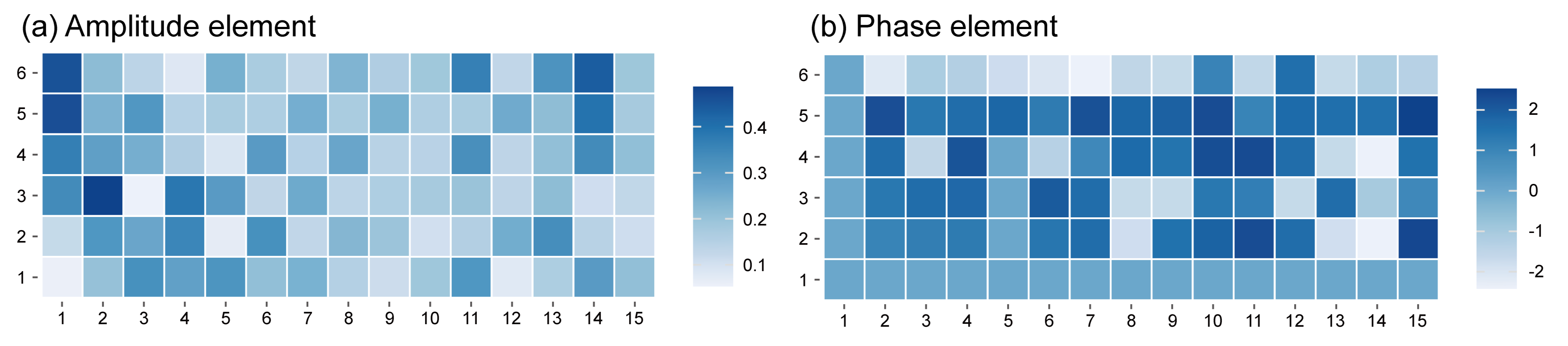}
		
		FIG. S3. \textbf{Experimental characterization of photonic chip.} \textbf{(a)} and \textbf{(b)} are experimentally reconstructed amplitude and phase elements in the Haar random unitary matrix.
		\label{s3}
	\end{figure}
	
	\clearpage
	
	\section{Time-of-flight module and timestamp method}
It is already known that the statistics of the photon counting satisfy Poisson distribution. Mathematically, we can simply prove that the exponential decay distribution shares the same nature with Poisson distribution if their parameter $\lambda$ is the same, just looking at the physical process from a different perspective. Each data point in exponential decay distribution represents the time interval between successive occurrences of events. Such process explains why the time interval is related to the event probability, thus we can retrieve the time information to reconstruct the output distribution, which is the main spirit of our timestamp method. 

For the experimental part, as shown in FIG.S4, the time-of-flight module records the electronic signals from the array of APDs, and the channels are colored in red while the blue color is used to mark the trigger channels. Multi-photon events are recorded when each single event occurs within the 2ns time window. We regard the triggers' registered time $T_i $ as the timestamp information of coincidence events and the probability of each combination $p_i$ can be calculated by $p_i=\frac{T_i^{-1}}{\sum_{i}T_i^{-1}}$.
		
	\begin{figure}[htbp]
		\centering
		\includegraphics[width=0.95\linewidth]{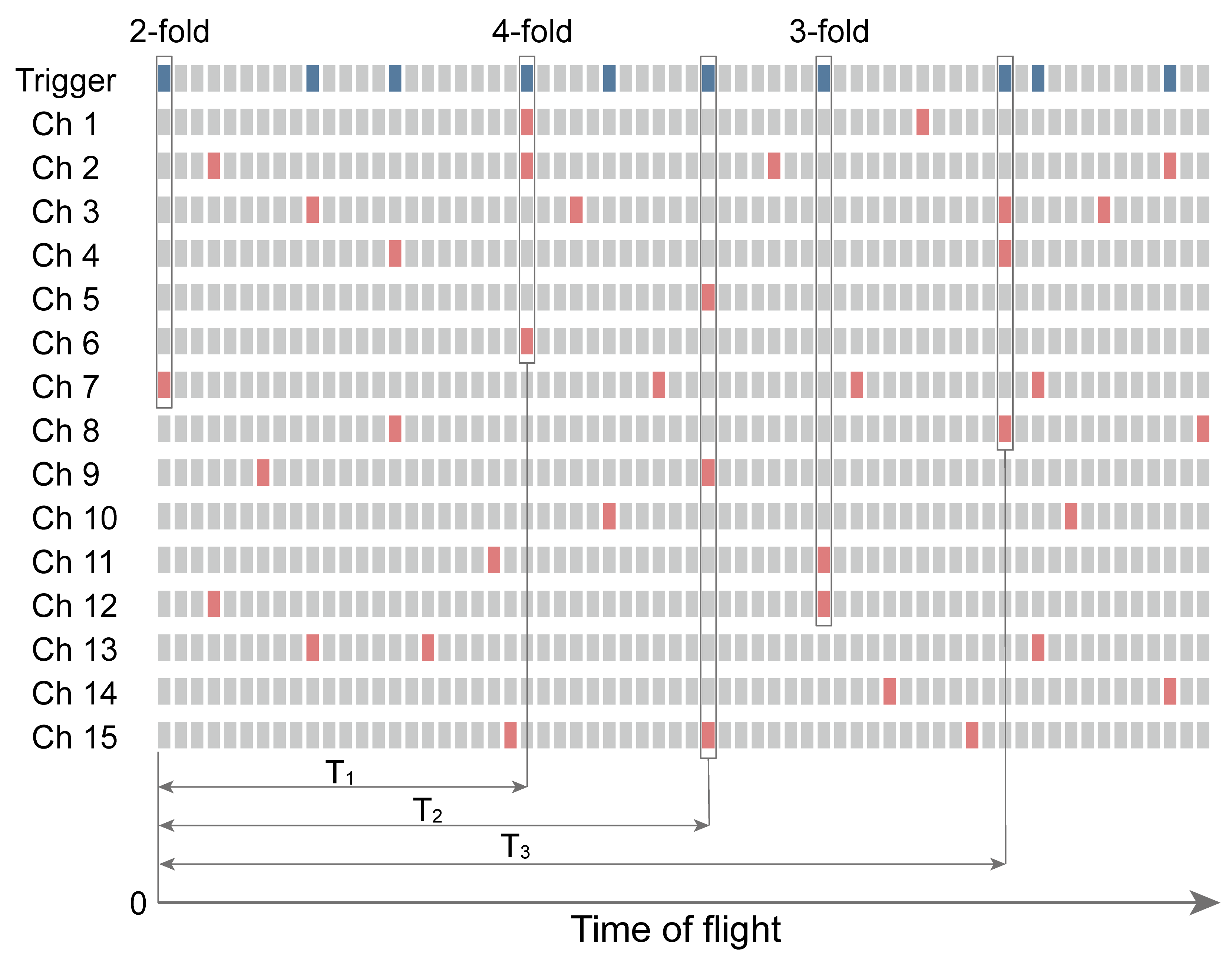}
		
		FIG. S4. \textbf{Time-of-flight module and timestamp method.} The time-of-flight module is addressed to the array of APDs and records the time information of the arrival electronic signals. Active channels are shown in red and blue blocks for trigger channel and signal channel. A 2ns coincidence time window is set to extract multi-photon events and the trigger registered time $T_i$ is the timestamp to mark different coincidence events.
		\label{s4}
	\end{figure}

	\section{Full photon distribution of 2-layer, 4-fold events}
The two-layer 4-fold coincidence events correspond to 8 photons scattershot in a 30-mode interferometer. We have listed the photon distributions in FIG.S5.
	
	\begin{figure}[htbp]
		\centering
		\includegraphics[width=0.75\linewidth]{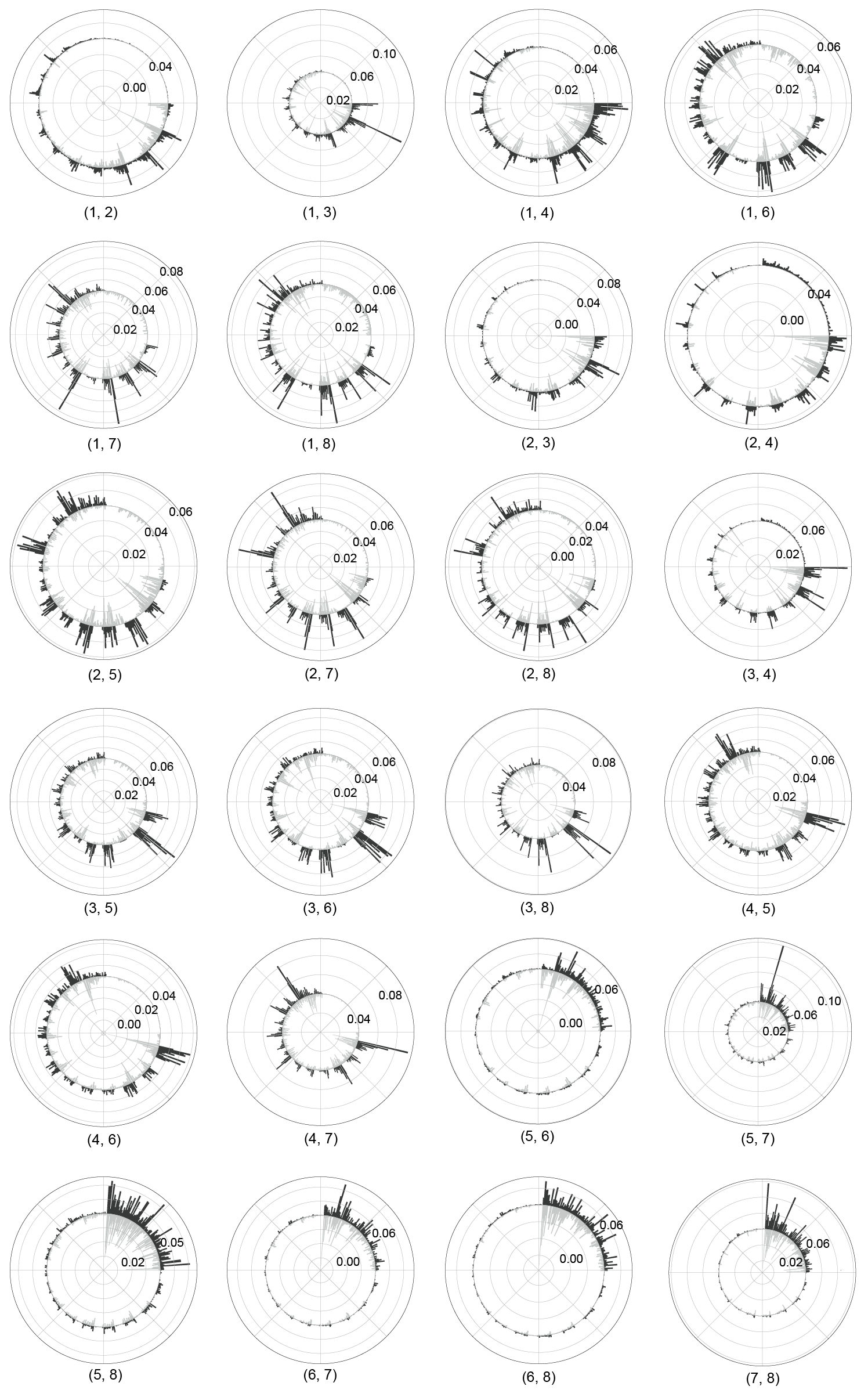}
		
		FIG. S5. \textbf{Full photon distribution of detectable 2-layer, 4-fold events.} The experimental data $p_i$ is plotted in dark grey while the theoretical data $t_i$ is given by light grey bars. The number at the bottom of the ring shows the input channel combinations.
		\label{s5}
	\end{figure}
	
	\clearpage
	
	\section{Time calibration for huge data processing} 
To deal with ultra huge layers, another issue we need to address is the time deviations. The repetition rate of the pulsed laser is not strictly equal to 80MHz, namely, the time interval between two pulses floats around 12.5ns. When the layer number increases, this float will result in a difference between the actual and theoretical time interval. So, if we still carry out Step II as the procedure of processing data in the small layers, we will regard multiple photon events by mistake. Additional time calibration is necessary, thus we numerically fit the relation to correct the time deviations . In order to improve the data processing efficiency, we also make use of the power of parallel computing on an Intel Xeon(R) Gold 6138 and 256GB RAM workstation.

We calculate this deviation dt and plot the relationship with layer intervals $L_{I}$, as shown in FIG.S6. We find that the relations are linear dependent and the relationship for the signal channels and the trigger signals share almost the same slope.
	
	\begin{figure}[htbp]
		\centering
		\includegraphics[width=0.9\linewidth]{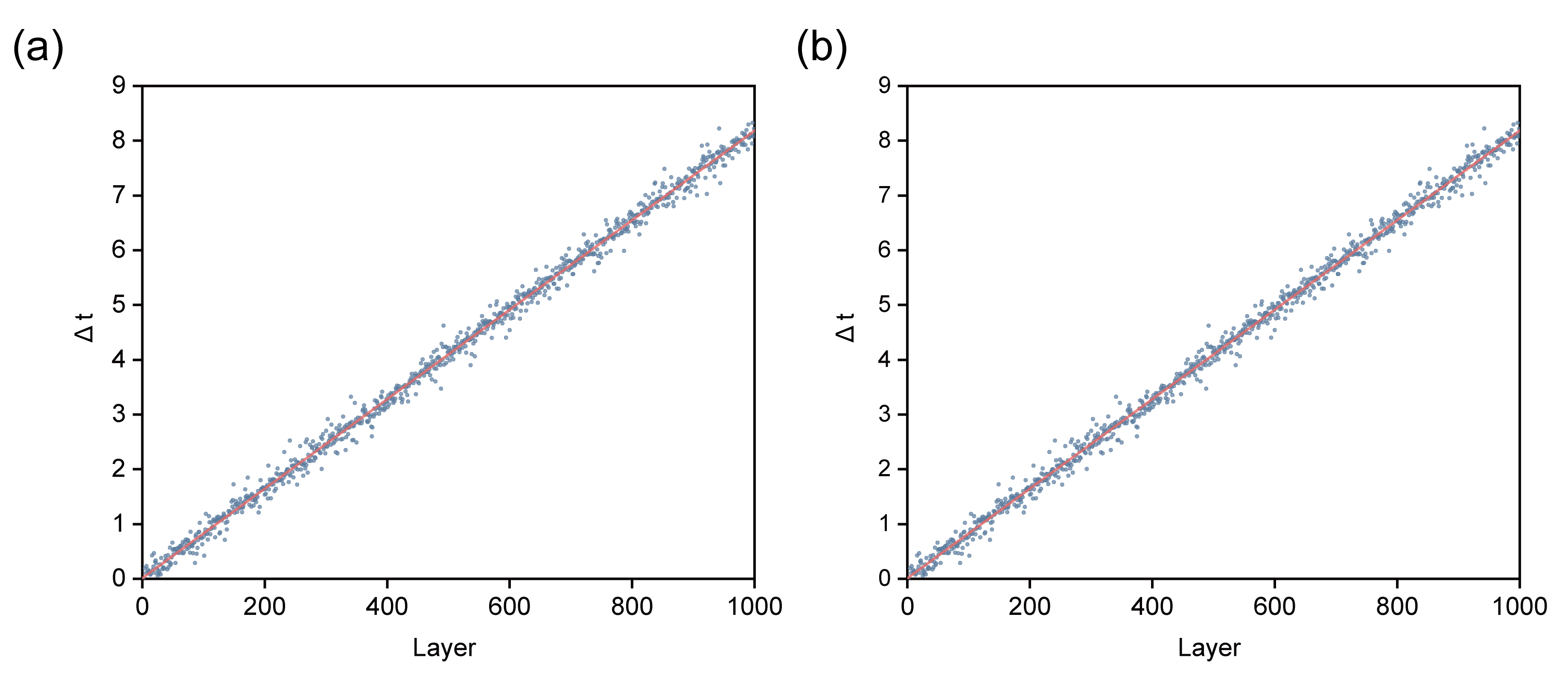}
		
		FIG. S6. \textbf{Fitting of time deviation for huge data processing.} \textbf{(a)} Fitting result of time deviation for the signal channels. The result is $dt = 0.008744\times L_{I}-0.2105$. The horizontal axis is section interval and the vertical axis is the time difference between the real time and theoretical time. When the section is larger than 700, the time deviation will be over 8ns, nearly half period. Theoretical result is layer separation $L_{I}\times$12.5ns. 
	\textbf{(b)} Fitting result of time deviation for the trigger channels. The result is $dt = 0.008358\times L_{I}-0.2764$ and the fitting result shows little dependence on signals or triggers.
		\label{s6}
	\end{figure}
	
	\clearpage


\begin{thebibliography}{99}
		\bibliographystyle{Science}
		
	\bibitem{Nielsen} M. A. Nielsen and I. L. Chuang, Quantum computation and quantum communication (Cambridge University Press, Cambridge, 2000).
	\bibitem{QC} T. D. Ladd, \textit{ et al.}, Quantum computers. \textit{Nature} \textbf{464}, 45 (2010).
	\bibitem{GisinRMP} N. Gisin, G. Ribordy, W. Tittel, and H. Zbinden, Quantum cryptography. \textit{Rev. Mod. Phys.} \textbf{74}, 145 (2002).
	\bibitem{error} S. J. Devitt, W. J. Munro, and K. Nemoto, Quantum error correction for beginners. \textit{Rep. Progr. Phys.} \textbf{76}, 076001 (2013).
	\bibitem{NISQ} J. Preskill, Quantum computing in the NISQ era and beyond. \textit{Quantum} \textbf{2}, 79 (2018).
	\bibitem{supremacy} F. Arute, \textit{et al.} Quantum supremacy using a programmable superconducting processor. \textit{Nature} \textbf{574}, 505 (2019).
	\bibitem{AA} S. Aaronson and A. Arkhipov, The computational complexity of linear optics. \textit{Theory Comput.} \textbf{9}, 143 (2013).	
	\bibitem{Vibronic} J. Huh, et al. Boson Sampling for Molecular Vibronic Spectra. \textit{Nat. Photonics} \textbf{9}, 615 (2015). 
	\bibitem{Spin} B. Peropadre, A. A. Guzik and J. J. G. Ripoll, Equivalence between Spin Hamiltonians and Boson Sampling. \textit{Phys. Rev. A} \textbf{95}, 032327 (2017).
	\bibitem{Graph1} J. M. Arrazola and T. R. Bromley, Using Gaussian Boson Sampling to Find Dense Subgraphs. \textit{Phys. Rev. Lett.} \textbf{121}, 030503 (2018).
	\bibitem{Graph2} K. Brádler, P.-L. Dallaire-Demers, P. Rebentrost, D. Su and C. Weedbrook, Gaussian Boson Sampling for Perfect Matchings of Arbitrary Graphs. \textit{Phys. Rev. A} \textbf{98}, 032310 (2018). 
	\bibitem{Graph3} M. Schuld, K. Brádler, R. Israel, D. Su and B. Gupt, Measuring the Similarity of Graphs with a Gaussian Boson Sampler. \textit{ Phys. Rev. A} \textbf{101}, 032314 (2020).
	%BS
	\bibitem{Oxford_2013} J. B. Spring, \textit{et al.}, Boson sampling on a photonic chip. \textit{Science} \textbf{339},6162 (2013).
	\bibitem{Vienna&Jena_2013} M. Tillmann, \textit{et al.}, Experimental boson sampling. \textit{ Nat. Photon.} \textbf{7}, 540 (2013).
	\bibitem{Roma&Milano_2013} A. Crespi, \textit{et al.}, Experimental boson sampling in arbitrary integrated photonic circuits. \textit{Nat. Photon.} \textbf{7}, 545 (2013).
	\bibitem{Roma&Milano_Birthday_2013} N. Spagnolo, \textit{et al.}, General rules for bosonic bunching in multimode interferometers. \textit{Phys. Rev. Lett.} \textbf{111}, 130503 (2013).
	\bibitem{Brisbane_2013} M. A. Broome, \textit{et al.}, Photonic boson sampling in a tunable circuit. \textit{Science} \textbf{339}, 6121 (2013).
	\bibitem{Roma&Milano_validation_2014} N. Spagnolo, \textit{et al.}, Experimental validation of photonic boson sampling. \textit{Nat. Photon.} \textbf{8}, 615 (2014).
	\bibitem{Bristol_Haar_random_2014} J. Carolan, \textit{et al.}, On the experimental verification of quantum complexity in linear optics. \textit{Nat. Photon.} \textbf{8}, 621 (2014).
	\bibitem{Bristol_Universal_2015} J. Carolan, \textit{et al.}, Universal linear optics. \textit{Science} \textbf{349}, 711 (2015).
	\bibitem{Vienna&Jena_2015} M. Tillmann, \textit{et al.}, Generalized multiphoton quantum interference. \textit{Phys. Rev. X} \textbf{5}, 041015 (2015).
	\bibitem{Roma&Milano_2016} A. Crespi, \textit{et al.}, Suppression law of quantum states in a 3D photonic fast Fourier transform chip. \textit{Nat. Commun.} \textbf{7}, 10469 (2016).
	\bibitem{Brisbane_QD_2017} J. C. Loredo, \textit{et al.}, Boson sampling with single-photon Fock states from a bright solid-state source. \textit{Phys. Rev. Lett.} \textbf{118}, 130503 (2017).
	\bibitem{Timebin} K. R. Motes, A. Gilchrist, J. P. Dowling, and P. P. Rohde, Scalable boson sampling with time-bin encoding using a loop-based architecture. \textit{Phys. Rev. Lett.} \textbf{113}, 120501 (2014).
	\bibitem{USTC_time_bin_2017} Y. He, \textit{et al.}, Time-bin-encoded boson sampling with a single-photon device. \textit{Phys. Rev. Lett.} \textbf{118}, 190501 (2017).
	\bibitem{USTC_QD_2017} H. Wang, \textit{et al.}, High-efficiency multiphoton boson sampling. \textit{Nat. Photon.} \textbf{11}, 361 (2017).
	\bibitem{USTC_Lossy_2018} H. Wang, \textit{et al.}, Toward scalable boson sampling with photon loss. \textit{Phys. Rev. Lett.} \textbf{120}, 230502 (2018).
	\bibitem{Oxford2019} B. A. Bell, G. S. Thekkadath, R. Ge, X. Cai, and I. A. Walmsley, Testing multi-photon interference on a silicon chip. \textit{Opt. Express} \textbf{27}, 35646 (2019).
	\bibitem{Collisionfree} J. Gao, \textit{et al.}, Experimental collision-free dominant boson sampling. http://arxiv.org/abs/1910.11320 (2019).
	\bibitem{USTC2019} H. Wang, \textit{et al.}, Boson sampling with 20 input photons and a 60-mode interferometer in a $10^{14}$ -dimensional Hilbert space. \textit{Phys. Rev. Lett.} \textbf{123}, 250503 (2019).
	\bibitem{Roma&Milano_Scattershot_2015} M. Bentivegna, \textit{et al.}, Experimental scattershot boson sampling. \textit{Sci. Adv.} \textbf{1}, e1400255 (2015).
	\bibitem{USTC_Scattershot_2018} H. S. Zhong, \textit{et al.}, 12-Photon entanglement and scalable scattershot boson sampling with optimal entangled-photon pairs from parametric down-conversion. \textit{Phys. Rev. Lett.} \textbf{121}, 250505 (2018).
	\bibitem{Bristol_Gaussian_2019} S. Paesani, \textit{et al.}, Generation and sampling of quantum states of light in a silicon chip. \textit{Nat. Phys.} \textbf{15}, 925 (2019).		
	\bibitem{Mem} P. Pfeiffer, I. L. Egusquiza, M. Di Ventra, M. Sanz, and E. Solano, Quantum memristors. \textit{Sci Rep} \textbf{6}, 29507 (2016). 
	\bibitem{Hybrid} X.-L. Pang, \textit{et al.}, A hybrid quantum memory enabled network at room temperature. \textit{Sci. Adv.} \textbf{6}, eaax1425 (2020).
	\bibitem{Beamlike} Y. H. Kim, Quantum interference with beamlike type-II spontaneous parametric down-conversion. \textit{Phys. Rev. A} \textbf{68}, 013804 (2003).
	\bibitem{Fab} K. M. Davis, K. Miura, N. Sugimoto, and K. Hirao, Writing waveguides in glass with a femtosecond laser. \textit{Opt. Lett.} \textbf{21}, 1729 (1996).
	\bibitem{Cha} A. Laing, J. L. O'Brien, Super-stable tomography of any linear optical device. http://arxiv.org/abs/1208.2868 (2012).
	\bibitem{QW_1D} A. Peruzzo, \textit{et al.} Quantum walks of correlated photons. \textit{Science} \textbf{329}, 1500 (2010).
	\bibitem{QW_2D} H. Tang, \textit{et al.}, Experimental two-dimensional quantum walk on a photonic chip. \textit{Sci. Adv.} \textbf{4}, eaat3174 (2018).
	\bibitem{LN} C. Wang, \textit{et al.}, Integrated lithium niobate electro-optic modulators operating at CMOS-compatible voltages. \textit{Nature} \textbf{562}, 101 (2018).
	\bibitem{timestamp} W.-H. Zhou, \textit{et al.}, Timestamp boson sampling. http://arxiv.org/abs/2009.03327 (2020).
	\bibitem{tianhe} J. Wu, \textit{et al.}, A benchmark test of boson sampling on Tianhe-2 supercomputer. \textit{Natl. Sci. Rev.} \textbf{5}, 715 (2018).
	\bibitem{Sunway} Y. Li, \textit{et al.}, Benchmarking 50-photon Gaussian boson sampling on the Sunway TaihuLight. http://arxiv.org/abs/2009.01177 (2020).
	\bibitem{SupremacyAA} A. W. Harrow, A. and Montanaro, Quantum computational supremacy. \textit{Nature} \textbf{549}, 203 (2017).
	\bibitem{Chalabi2019} H. Chalabi, \textit{et al.}, Synthetic gauge field for two-dimensional time-multiplexed quantum random walks. \textit{Phys. Rev. Lett.} \textbf{123}, 150503 (2019).

	\end{thebibliography}
\end{document}